# A systematic review of structural equation modeling in augmented reality applications

**Vinh The Nguyen[1], Chuyen Thi Hong Nguyen[2]**

[1]Faculty of Information Technology, TNU-University of Information and Communication Technology, Thai Nguyen, Vietnam
[2]Faculty of Primary Education, Thai Nguyen University of Education, Thai Nguyen, Vietnam

## Article Info

## ABSTRACT

The purpose of this study is to present a comprehensive review of the use of structural equation modeling (SEM) in augmented reality (AR) studies in the context of the COVID-19 pandemic. IEEE Xplore Scopus, Wiley Online Library, Emerald Insight, and ScienceDirect are the main five data sources for data collection from Jan 2020 to May 2021. The preferred reporting items for systematic reviews and meta-analyses (PRISMA) approach was used to conduct the analysis. At the final stage, 53 relevant publications were included for analysis. Variables such as the number of participants in the study, original or derived hypothesized model, latent variables, direct/indirect contact with users, country, limitation/suggestion, and keywords were extracted. The results showed that a variety of external factors were used to construct the SEM models rather than using the parsimonious ones. The reports showed a fair balance between the direct and indirect methods to contact participants. Despite the COVID-19 pandemic, few publications addressed the issue of data collection and evaluation methods, whereas video demonstrations of the augmented reality (AR) apps were utilized. The current work influences new AR researchers who are searching for a theory-based research model in their studies.

*Corresponding Author:*

Vinh The Nguyen
Faculty of Information Technology, TNU-University of Information and Communication Technology
Z115 Street, Quyet Thang Commune, Thai Nguyen, Vietnam
Email: vinhnt@ictu.edu.vn

## 1. INTRODUCTION

Augmented reality (AR) is a technology that has attracted a lot of attention in various domains [1]-[3]. Unlike virtual reality (VR) which allows users to be totally immersed in a virtual environment, AR enriches the real world with virtual artifacts [4]. The primary value of AR is that it allows digital objects to be blended more seamlessly into a person's perception of the real world than simply displaying data on a screen. Market research [5] anticipates that AR's market will reach USD 88.4 billion, growing 31.5% from 2021 to 2026. In addition, in response to the COVID-19 pandemic, more companies and organizations have adopted remote work and are utilizing augmented reality technology [6]. What that means is that a huge number of AR applications are being developed, especially in electrical engineering and computer science [1]-[3], [7], [8].

Assessment is one of the key factors in ensuring the success of an AR application, especially when it is involved with end-users. However, literature work reported that only a few studies afforded time for this type of evaluation (only 8% of published papers) [9]. One plausible explanation was that AR researchers/developers had to devote their time to solving technical issues [10]. Moreover, the lack of methods or theory-driven research on evaluating AR apps, considering end users' involvement, contributed to the scarcity of AR evaluation [11]. In addition, after the COVID-19 outbreak, many conferences (e.g., ISMAR) encouraged researchers to find alternative means of evaluating AR apps rather than canceling the submissions due to social distancing. There has been no study addressing this issue so far, thus it remains a gap in the literature. To close this gap, this paper-based on prior AR studies–provided an overview of theory-based methods that can effectively be used for AR assessment. Among many other end-user evaluation methods, the scope of the current study focused on structural equation modeling (SEM), a model commonly used in behavioral science. SEM is a comprehensive statistical method that examines relationships between observed and latent factors [12]. It has been widely used in confirmatory factory analysis in many topics and fields [13]-[15].

A number of review studies on SEM applications have been conducted in various research domains, including ecology [16], social science [17], psychological research [18], and strategic management [19]. It indicated that a review study would be valuable for new researchers to quickly acquire knowledge in the field effectively. Yet, it also implies that it would be important to look at SEM from AR's perspective since AR is one of the emerging trends in the digital transformation era. However, there is no study of SEM for AR applications other than previously mentioned review studies. Thus, the current research is unique on its own by the AR's topic and the outcomes of this study can be used as a referencguidene for researchers in similar studies, particularly in electrical engineering and computer science. More specifically, the present study tries to answer to following research questions: i) What are the preferred theory-driven models being used in prior AR studies amid the COVID-19 pandemic? ii) What are the dimensions or variables being investigated by AR researchers so far? iii) How do researchers of prior AR studies communicate with end-users for evaluation? vi) How many participants are typically involved in a study? Would this number still be considered appropriate from the literature? v) What are the main drawbacks of tR studies? Do they suffer from the COVID-19 pandemic?

## 2.    METHOD

This study involves a review of SEM in AR applications; thus, the preferred reporting items for systematic reviews and meta-analyses (PRISMA) statement was applied [20]. The PRISMA statement aims to assist scholars in improving the reporting of scientific reviews and meta-analyses. It is an evidence-based minimum set of elements for systematic review reports that are intended to assist systematic reviewers in clearly explaining why the review was conducted and what the authors performed. It has previously been used to target comparable research objectives [21], [22].

### 2.1.  Source selection

IEEE Xplore, Scopus, Wiley Online Library, Emerald Insight, and ScienceDirect databases were used to build the corpus, encompassing titles, abstracts, and keywords. These five databases are regarded as essential and dependable sources of high-quality articles in the fields of computer science and engineering [21], [23]. Although, some other indexing databases are available (i.e., Scholar) but they are out of scope in the current study.

### 2.2.  Search criteria

To add articles to our corpus, both of the following related criteria need to be fulfilled, i) Structural equation modeling search term: at least one SEM-related term must appear in an article's title, abstract, or author keywords (i.e., structural equation modeling, SEM, planned behavior, theory of planned behaviour (TPB), motivational model, Michaelis–Menten (MM), reasoned action, theory of reasoned action (TRA), social cognitive, SCT, diffusion of innovation, IDT); and ii) Augmented Reality search term: terms include augmented reality, AR. Using the aforementioned criteria, 16 articles were discovered in IEEE Xplore, 107 articles in Scopus, 197 papers in Wiley Online Library, 68 papers in Emerald Insight, and 695 papers in ScienceDirect. The corpus was collected between June 3, 2021, and June 12, 2021.

### 2.3.  Eligibility assessment for the final analysis corpus

To determine the acceptability of the obtained papers, the first researcher personally reviewed the entry criteria mentioned below by reviewing the titles and abstracts of the obtained publications. When a clear judgment could not be reached, other aspects of the publication, particularly the method and data acquisition descriptions, were discussed in conjunction with the second author. Only items that meet the following criteria are retained in the corpus: i) Peer-reviewed: The paper was peer-reviewed in the two indexing databases. This is due to the trustworthiness of peer-reviewed journals and the rigorous peer-review processes, only articles in these databases are considered for this review; ii) Topic relevant: The topic of an article is pertinent to the applications of SEM in AR; iii) Language: Publication was reported in English; and vi) Duration: Paper was published between Jan 2020 and May 2021.

If the article meets any of the following criteria, it will be excluded from the corpus: i) Books and cover page, abstract only, poster; ii) The paper was not written in English; iii) Application of SEM is not for AR; and vi) Paper was published before Jan 2020 and after May 2021.

Figure 1 depicts the flow of information through the different phases of the systematic review utilizing PRISMA approach. 1,083 records were found in all data sources. Duplications were removed based on the titles. Each paper was screened individually to remove items that are out of scope. Then 230 records were excluded. As such, 309 candidates left for full-text retrieval. Of these remaining items, 9 records cannot be retrieved due to access restrictions. The authors examined each report for eligibility and removed 247 studies. In the end, 53 items were included in this research. The remaining papers were examined individually to extract interesting variables such as the number of participants, original or derived hypothesized model, latent variables, direct/indirect contact with the user, country of origin, limitation/suggestion (if any), and keywords.

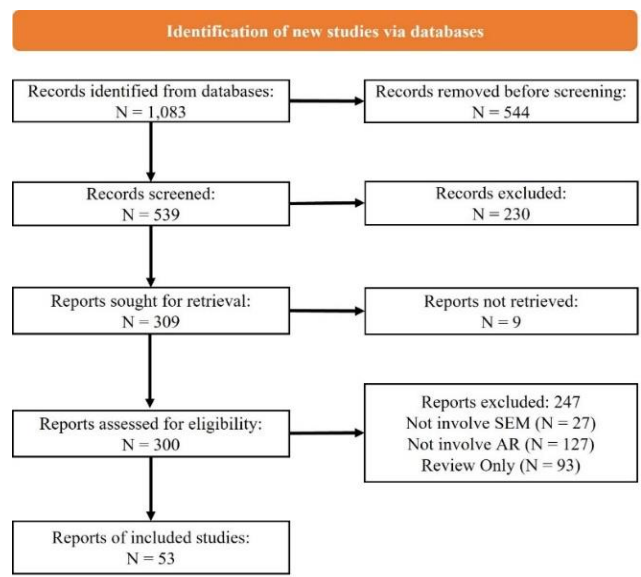

Figure 1. The flow diagram represents the movement of information through the various stages of a systematic review

## 2.4. Data coding and analysis

To extract the data, all articles were loaded into NVivo software, and a coding scheme was created. NVivo is a program that facilitates qualitative analytical method research. This tool enables researchers to organize, analyze and explore unstructured or qualitative data, including interviews, reviews, articles, social media, and web content. Codes included authors, journal name, year of publication, countries of authorship, title, abstract, author keywords, method, objectives, findings and limitations on how SEM was used.

## 3. RESULTS AND DISCUSSION
### 3.1. What are the preferred theory-based driven models being used in prior AR studies amid the COVID-19 pandemic?

Figure 2 depicts the distribution of papers over hypothesized models. Most publications fall into the SEM category (accounted for 58.49%), followed by eTAM and TAM with 20.76% and 11.32% respectively. Although the UTAUT model was developed recently, the result shows less popularity of adopting this model (only 3.77%), which is the same as the SOR model.

Technology acceptance model (TAM): originally developed by Davis [24], TAM is known as a theory of information systems that describes how consumers come to accept and use technology. Real system usage is the point at which people interact with technology. People utilize technology because of their behavioral intentions. In this survey, 6 articles (11.32%) used original TAM for their research.

Extended technology acceptance model (eTAM): In this category, 11 publications (20.75%) extended TAM with external variables such as perceived task-technology fit [25]-[28]–which asserted that the technology must be utilized and a good fit with the tasks it supports to have positive impacts on individual performance, perceived visual design/appeal [25]-[31] which assumed that beauty is important, and it impacts decisions that should not be influenced by aesthetics, perceived enjoyment [32]-[35]-which refers to the hedonic value of new technology and expresses how pleasurable a person finds its use.

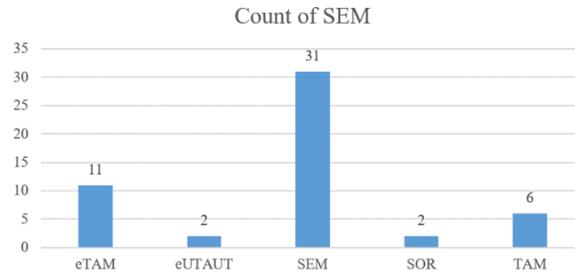

Figure 2. Models distribution across prior studies

The unified theory of acceptance and use of technology (UTAUT): Venkatesh *et al*. [36] developed the UTAUT after reviewing and consolidating the components of eight previous models used to describe information system user behavior. In this review, several external variables were incorporated into the existing UTAUT model (eUTAUT) such as innovativeness, reward, trust, enjoyment, hedonic motivation, habit, and gamification [37], [38].

Stimulus-organism-response (SOR): Mehrabian-stimulus Russell's model [39] depicts the occurrence of a person's response to environmental stimuli. Qin *et al*. [40] decomposed stimulus into two external factors (i.e., Interactivity, Virtuality), Organism into 4 variables (i.e., Hedonic, Utilitarian, Informativeness, and Ease of Use), and Response into 2 factors including Attitude and Behavioral Intention. Similarly in the scope of this review, Qin *et al*. [40] also included (critical mass, social interaction, information timelines, content richness) into stimulus, (attachment, conformity) into Organism, and (visiting intention, continue intention) into Response.

Structural equation modeling (SEM): This category contains the largest portion of the papers included in our investigation (58.49%). Authors in this group mainly adapted constructs, measures in the literature to form hypothesis. As such, PLS-SEM was utilized as an analytical method to conduct confirmatory factor analysis and path analysis. Confirmatory factor analysis, which originates in psychometrics, aims to quantify underlying psychological characteristics such as attitude and satisfaction. Path analysis, on the other hand, has its origins in biometrics and is intended to discover the causal link between variables by drawing a path diagram [41].

### 3.2.  What are dimensions or variables being investigated by AR researchers so far?

Figure 3 depicts 77 unique constructs/latent variables from hypothesized models. There are 184 unique constructs found in this study. Behavioral intention, usefulness, ease of use, attitude, user behavior, and enjoyment are the most frequent items used in the hypothesized models.

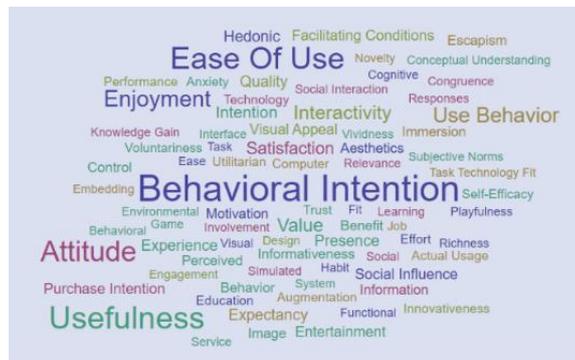

Figure 3. Wordcloud depicts 77 unique constructs from all hypothesized models

Figure 4 captures the top 14 dominant keywords in the collection of papers in this study. Aside from "augmented reality", TAM is the most popular term that the authors used for indexing their papers. In total, this study extracted 319 keywords with 230 unique terms, indicating that there is a high variation of topics/techniques used. However, in terms of their broad contents, the major theme of these collected papers can be categorized as the "social marketing" theme as they were mainly focused on "Intention to Purchase" or "Intention to Visit".

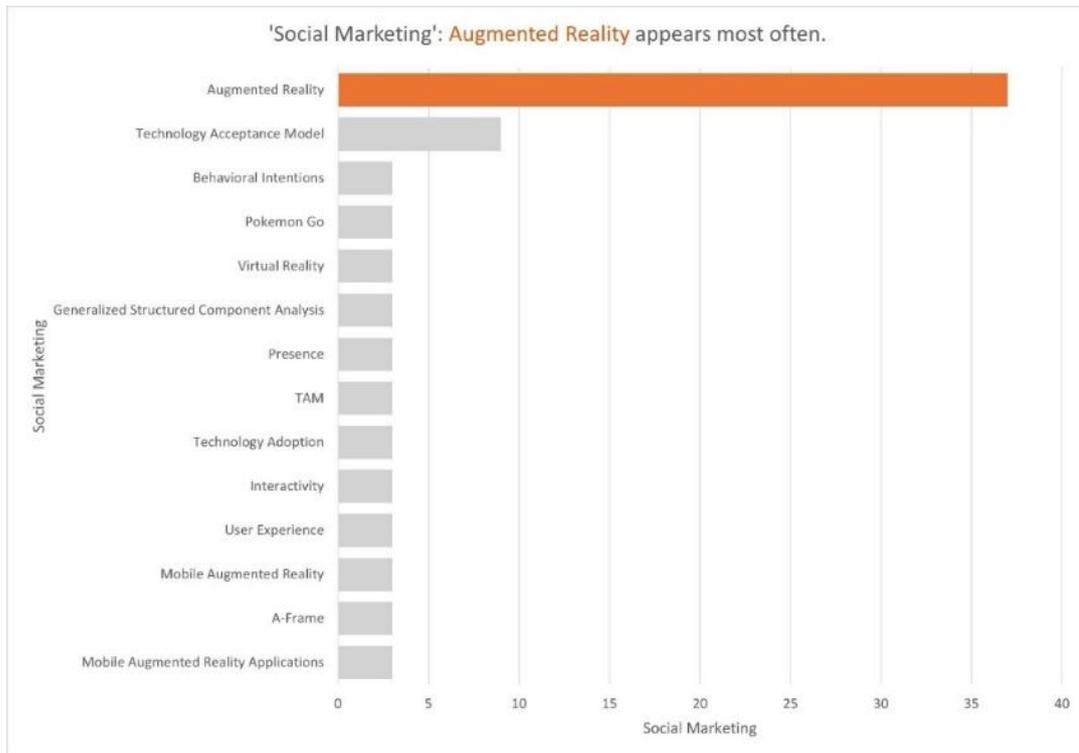

Figure 4. Frequency of keywords extracted from publications

### 3.3. How do researchers of prior AR studies communicate with end-users for evaluation?

Table 1 reports the communication channels used to gather data from respondents. Results showed that there is a fair balance between the direct (45.28%) and indirect (50.94%) methods. Here, the indirect method means that the research teams did not contact participants directly (e.g., lab setting, or field study). Instead, they contact users via online channels (e.g., social network, email, discussion group). On the other hand, the direct method requires subjects to be at the site of the study for the experiment.

Table 1. Communication channels to collect data from respondents

| Communication channel | Count | Percentage |
|---|---|---|
| Indirect | 27 | 50.94 |
| Direct | 24 | 45.28 |
| Direct and Indirect | 2 | 3.77 |
| Total | 53 | 100 |

Figure 5 depicts the spatial locations of authors researching AR utilizing the SEM method across the globe. It can be observed that most publications were conducted in the United States although this country was suffered heavily from the COVID-19 pandemic. However, 8 out of 10 papers utilized the indirect research method to recruit and gather data, meaning that the study was conducted remotely, and opinions were collected through online tools.

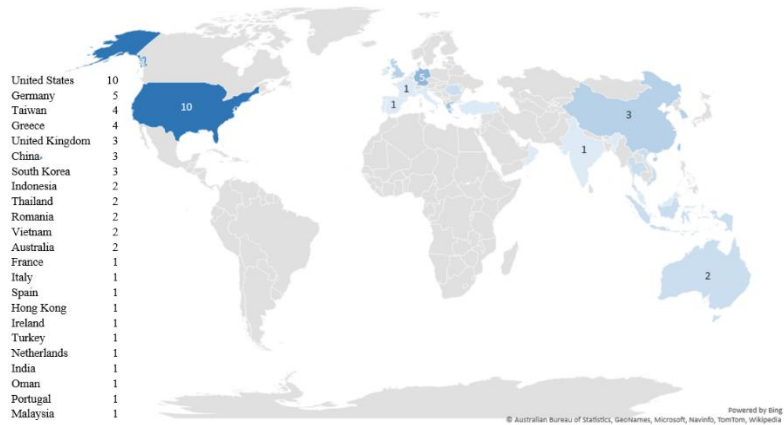

Figure 5. Spatial locations of authors researching on AR utilizing SEM in 2020-2021

## 3.4. How many participants are typically involved in a study? Would this number still be considered appropriate from the literature?

Figure 6 shows the distribution of sample size across peer-reviewed papers. The whisker plot indicates that on average the sample size (the number of participants) who took part in the studies was approximately 300 subjects considering 4 extreme values (or outliers). The minimum sample size is 9 and the maximum is 1,566. The median indicates that most papers recruited around 200 users for their studies. When the four extreme values were not considered, the average sample size for direct communication with participant was 142 (median=113, range=340, min=24, max=364), and indirect method was 286 (median=302, range=710, min=9, max=719).

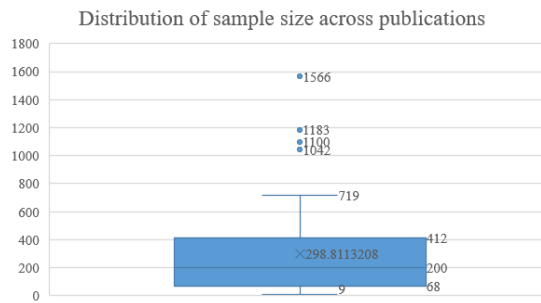

Figure 6. Distribution of sample size in the peer-reviewed papers

Sample size is a debating subject in the literature. As such, the determination of sample size varies from study to study. Some researchers advocate a minimum sample size of 100–200 per a study, an acceptable sample size can range between 300 and 500, or with criteria such as acceptable of five cases per free parameter, moderate of ten cases per free parameter [12], and ideal of 20 instances per free parameter in the model. Kock and Hadaya [42] proposed a technique for determining an adequate sample size based on "inverse square root" and "gamma-exponential" approaches which were adapted by Nikhashemi *et al.* [43] included in this study. To some extent, Figure 6 reflects the balance of sample size recommendation in the literature. Interestingly, the median sample size calculated in this study (Median=200) was aligned with the findings based on reviews of studies in different research areas, including operations management, education and psychology.

## 3.5. What are the main drawbacks of the AR studies? Do they suffer from the COVID-19 pandemic

Table 2 reports the frequency of limitations addressed by authors in the collected publications. The most common flaw that needs to be examined further in future studies is the failure to incorporate additional external components (39.62%) in the postulated model, followed by convenience sampling (35.85%), multi-level analysis (32.08%) and limited to one region (30.19%). In terms of convenience sampling drawback, many authors acknowledged that they used the non-probability method to acquire sample data through their networks of interest. As such, their reports/findings cannot be generalized to the population.

Table 2. Frequency of limitations addressed by the authors in the collected publications

| Limitations | References |
| --- | --- |
| Not consider other factors (21) | [25], [26], [30], [31], [32], [38], [43], [44], [45], [46], [47], [48], [49], [50], [51], [52], [53], [54], [55], [56], [57] |
| Convenience sampling (19) | [32], [35], [37], [40], [43], [45], [46], [50], [52], [53], [55], [57], [58], [59], [60], [61], [62], [63], [64] |
| Multi levels analysis (17) | [25], [37], [44], [45], [46], [48], [49], [51], [55], [56], [59], [62], [65], [66], [67], [68], [69] |
| Limited to one region (16) | [30], [32], [33], [37], [38], [46], [47], [48], [49], [54], [56], [58], [59], [63], [68], [70] |
| Tailored to a specific AR product (14) | [45], [46], [47], [52], [53], [54], [57], [61], [62], [64], [65], [67], [68], [70] |
| Small Sample Size (10) | [30], [32], [33], [38], [40], [47], [50], [54], [60], [71] |
| Short term effect (10) | [29], [31], [38], [43], [45], [58], [63], [65], [69], [72] |
| Not specified (9) | [34], [41], [50], [73], [74], [75], [76], [77], [78] |
| Only Intention Model (6) | [31], [51], [52], [56], [58], [79] |
| Lack of AR features (6) | [25], [29], [32], [48], [63], [71] |
| Lack of functions (4) | [25], [26], [29], [32] |
| Self-Administered Survey (3) | [58], [66], [79] |
| Use Videos for demonstrations (3) | [25], [26], [65] |
| Technical challenges (2) | [27], [28] |
| Standardized tools (2) | [29], [52] |
| Single Analysis technique (2) | [33], [48] |
| Lab setting (2) | [55], [64] |
| Not consider privacy concerns (2) | [25], [60] |
| Others (8) | [25], [26], [29], [32], [56], [59], [58], [70] |

Along with convenience sampling, limited study to one region is another shortcoming that is often mentioned with non-probability method limitation. Unlike convenience sampling drawback that subjects may come from different parts of the world, the regional issue was arising where the study was intentionally designed for a specific region through a case study or in the lab setting [55], [64]. A large portion of the published work was carried out with the help of pre-existing AR products. This evaluation includes examples such as IKEA Place, YouCam Makeup, and Pokémon Go. Participants were asked if they had any experience with these AR apps, and if so, they were encouraged to take part in the survey. Furthermore, the authors' capacity to extend the study to additional products/services was limited because they did not have control or flexibility over the AR apps.

The results show that though the sample size was a sufficiently addressed problem by the researchers, the proportion of this limitation was just 18.87%. Without considering publications that did not report limitations in their work (i.e., not specified (9)), 77.27% (34/44 papers) of the research group justified their sample size using an analytical tool/method, a sample size recommendation in the literature, and the use of PLS-SEM, which can work with small sample sizes. As a result, sample estimation was deemed sufficient. Another issue worth mentioning is the short-term effect addressed by 10 author groups (18.87%). The short-term impact was explained by the fact that the experiments were only conducted for a limited period. As a result, the theorized models can only explain variables impacting user behavior at that point in time. The authors emphasized that because technology has evolved drastically over the years, the question of whether their proposed models stand up remained unresolved. In addition, people's perspectives shift throughout time as they gain experience [36], as a consequence, long-term research was suggested to validate the models.

In terms of the indirect method to conduct an experiment with users, four studies administered their AR applications through video demonstrations [25], [26], [31], [65]. In this regard, instead of asking participants to download or use the AR apps directly, the authors created videos demonstrating the features of their studied AR apps. Based on the evidence of previous studies using video depictions of AR prototypes [80], [81], these authors argued that the technology itself was not available for participants to interact with at the time, and the purpose of the hypothesized models was to examine the influential factors that affect behavioral intention before releasing the actual AR product to the market. As such in this category, studies in [26], [29], [52] recommended that there is a need to have a tool or new evaluation method to overcome the current issue.

In summary, compared with previous studies [16]-[19], this study has some similarities and differences as: First, it is the selection of model, our report also shows similar results, that is, many different types of models and variables are applied to the research. There has not yet been a general consensus set to guide new researchers to follow. The difference is that the variables in this study revolve around technology rather than ecology, social science, psychology, and management. Second is the issue of limitations. While similar studies only listed restrictions that exist in articles, our study quantified these limitations by specific numbers and arranges them in descending order. As such, interested researchers can rely on it to cover the information more broadly. The third consideration is the study's time span. This investigation was carried out in the context of digital transformation and the influence of COVID-19. Many new factors emerge and exert effect that have received little consideration in prior research (see Figure 3). Summarizing these factors will help researchers have more options instead of reading different articles. And finally, by synthesizing how the experiments were carried out during the pandemic, not only new researchers can adapt prior evaluation approach in the current situations but also improve them in the subsequent studies.

## 4.  CONCLUSION

This paper presented a systematic review of the use of SEM in AR studies during the COVID-19 pandemic. The PRISMA model was adapted as a guideline for doing the research. Five data sources were used for data retrieval. After a series of preprocessing steps, 53 publications were included in the study. The results showed that authors use a variety of external factors to form the generative hypothesized models (SEM), followed by the extension of TAM. The diversity of external factors indicated that there is no consensus among AR scholars for using common factors influencing AR adoption, thus opening a huge potential research gap for the AR community. Interestingly, United States was the most active country in conducting AR studies during the Covid-19 pandemic, however 80% of its studies were conducted through indirect communication channels. Hence, they were not affected by the pandemic. A large portion of AR studies focused on understanding factors influencing user behavioral toward using third-party AR apps. As such, participants were required to download and use the apps then answer the survey questionnaires. Sample size, in this regard, cannot be excused due to social distancing. Only few studies examined user behavioral through developed AR apps and the corresponding authors suggested that there is a need to have an alternative approach to conduct user study rather than the traditional face-to-face fashion. Watching two separate videos (one with AR and one without AR) was currently be used as an alternative method to alleviate the issue but not a plausible approach in the long run. Therefore, this research gap remains open and needs to be addressed in further studies. Thus, the outcomes of this study can be used as a reference guideline for researchers in similar studies where there is a lack of theoretical framework for assessment, particular in electrical engineering and computer science.

## ACKNOWLEDGEMENTS

This research is supported by project T2022-07-09 undertaken at the TNU–University of Information and Communication Technology, Thai Nguyen, Vietnam.

## BIOGRAPHIES OF AUTHORS

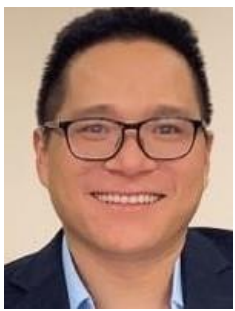

**Dr. Vinh The Nguyen** 🆔 🔗 SC P is currently a lecturer at the Faculty of Information Technology, University of Information and Communication Technology. He is also a senior visiting lecturer at FPT University Greenwich, Hanoi branch. He graduated with a master's degree in information systems management from Oklahoma State University, USA (under scholarship 322). He completed his PhD program under Project 911 in 2020 at Texas Tech University, USA. His main research interests are Computer Vision, Computer Visualization, and Computer in Human Behavior. He has authored or coauthored more than 35 publications with 10 H-index and more than 250 citations. He can be contacted at email: vinhnt@ictu.edu.vn.

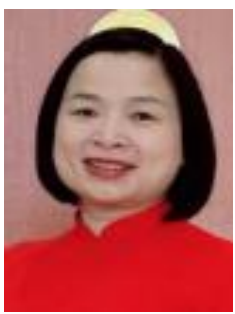

**Chuyen Thi Hong Nguyen** 🆔 🔗 SC P is currently a lecturer at the Faculty of Primary Education, Thai Nguyen University of Education, Vietnam. She graduated with a master's degree in Theory and History of Education from Hanoi University of Education, Vietnam (2008). She completed her PhD program in 2016 at The Vietnam Institute of educational Sciences, Vietnam. Her main research interests are method teaching, assessment in primary education, computational thinking, learning style, and augmented reality in education. She can be contacted at email: chuyennh@tnue.edu.vn.